# An Artificial Neural Net approach to forecast the population of India

**Goutami Bandyopadhyay and Surajit Chattopadhyay***

1/19 Dover Place

Kolkata-700 019

West Bengal

India

* E-mail: surajit_2008@yahoo.co.in

## Abstract

Present paper endeavors to forecast the population of India through Artificial Neural Network. A non-linear Artificial Neural Net model has been developed and the prediction has been found to be sufficiently accurate. It has been found that the model is performing more efficiently in predicting female population than the male population. Results have been presented graphically.

**Key words: India, male population, female population, age group, total fertility rate, life expectancy, prediction, chaos, Artificial Neural Network**

## Introduction

Forecasting the trends in human population is a complex problem. There are a number of uncertainties associated with the population of any country. If population of a country were considered as a *predictand*, then it would be found that there are a number of feasible predictors. Some predictors may be highly correlated with the *predictand* and some may not be so highly correlated. Furthermore, type and number of predictors vary from country to country. Some social or political changes can also bring in some change in the patterns of the predictors. For such reasons, the authors of this paper think that population of a country should be considered as a chaotic system. Here, some small change in one of the conditioning factors may bring in a sizeable change in the population at a given point of time. Traditionally, statistical approaches are used to forecast the demography of a country. Statistical demographic prediction agencies produce two or more forecasts of fertility or mortality (or both) to cope with the huge uncertainty associated with the demography of a country (Keilman et al, 2002). But, traditional statistical approaches are not very suitable for predicting a chaotic system like population. Firstly, statistical approaches make some assumptions, which are sometimes found unrealistic (Keilman et al, 2002). Secondly, statistical population forecast procedures cannot deal with the intrinsic chaos.

Alho and Spencer (1985), De Beer (1997), Alho (1997), Keilman et al (2002) discussed different sources of uncertainty in the prediction of population in a given country. Alho (1990) discussed stochastic methods of population forecasting. Ruggles (1992) explained the role of migration, marriage, and mortality as correcting sources of bias. Keilman (2002) discussed some limitations of traditional approaches in statistical population prediction and explained how various correlation structures associated with demography vary from country to country. Jonker and van der Vaart (2005) discussed some interesting sources of uncertainty in the demographic research and estimated mortality using the theory of conditional probability and distribution functions. Concept of spline interpolation has been explored by Smith et al (2004) to view the cumulative death rate of Australian females.

All the earlier studies are based either upon statistical procedures or upon conventional mathematical techniques. But, to analyze the population data and to predict their trend in a given country requires the application of some technique that can deal with the intrinsic chaotic nature. Introduction of Artificial Neural Network (ANN) has immensely empowered the forecasting techniques of intricate systems since the last few decades. Electrical load forecasting (Chaturvedi et al, 2003; Guangxi and Manli, 1993), meteorological forecasting (Smith et al 2005; Jain 2003), environmental prediction (Perez et al, 2000) etc have been enormously improved by the application of Artificial Neural Network. But, a literature survey shows that despite immense intricacy, population has not been the area of application of ANN. Yuan and Hsu (2006) implemented ANN in analyzing the impact of population upon industrial structure.

Present paper aims to implement ANN in predicting the population of India using ANN with Delta Learning algorithm. Detailed implementation procedure and the outcomes are presented in the subsequent sections.

## Artificial Neural Network-a brief review

Artificial Neural Network (ANN) is useful in the situations, where underlying processes / relationships may display chaotic properties. ANN does not require any prior knowledge of the system under consideration and are well suited to model dynamic systems on a real-time basis (Maqsood *et al.*, 2002). The perception of Artificial Neural Network instigated from the endeavor to build up a mathematical replica capable of recognizing intricate patterns on the same line as biological neurons works. The basic attributes of neural networks may be divided into the architecture and the functional properties or neurodynamics. Architecture defines the network structure, that is, the number of artificial neurons in the network and their interconnectivity (for details see Kartalopoulos, 2000). Learning an ANN is highly important and is undergoing intense research in both biological and artificial neural network. In the present paper a Generalized Delta Rule also named as Backpropagation learning is adopted to train Feed Forward Neural Network developed on the basis of various population related data.

The generalized delta rule (Kartalopoulos, 2000; Perez et al, 2000) is one of the most commonly used learning rules. For a given input vector, the output vector is compared to the correct result. If the difference is zero, no learning takes place; otherwise, the weights are adjusted to reduce this difference. The learning is done by least-square-error minimization. The least- square -error ($E$) between the target output ($T$) and actual output ($O$) is given by:

$$E = \frac{1}{2}(T_i - O_i)^2 = \frac{1}{2}[T_i - f(w_i x_i)]^2 \qquad \ldots(1)$$

Where, $w_i \Rightarrow$ The weight matrix associated with $i^{th}$ neuron

$x_i \Rightarrow$ Input of the $i^{th}$ neuron

$O_i \Rightarrow$ Actual output of the $i^{th}$ neuron

$T_i \Rightarrow$ Target output of the $i^{th}$ neuron

The activation function $f(x)$ is taken as the sigmoid function

$$f(x) = \frac{1}{1 + e^{-x}} \qquad \ldots \qquad \ldots \qquad \ldots \qquad (2)$$

The error gradient vector is

$$\nabla E = -(T_i - O_i) f'(w_i x_i) x_i \qquad \ldots \qquad \ldots \qquad \ldots \qquad (3)$$

Since minimization of the error is the objective, the weight is updated as

$$\Delta w_i = -\mu \nabla E \qquad \ldots \qquad \ldots \qquad \ldots \qquad (4)$$

Where $\mu$ is a positive constant.

Hence, $\Delta w_i$ becomes;

$$\Delta w_i = \mu (T_i - O_i) f'(w_i x_i) x_i \qquad \ldots \qquad \ldots \qquad \ldots \qquad (5)$$

Applying discrete mathematics, the weight vector is updated as:

$$w_i(k+1) = w_i(k) + \mu (T_i - O_i) f'(w_i x_i) x_i \qquad \ldots \qquad \ldots \qquad (6)$$

## Problem Definition and Methodology

Present paper develops a Multilyer Perceptron Model with Generalized Delta Learning to predict the population of India. Each year India adds more people to the world's population than any other country. Indian population trend is complex attributable to various reasons. The reasons can be summarized as follows:

- Fertility rate has declined but the number of women in their reproductive age has increased rapidly
- States of India vary significantly with respect to fertility, mortality and contraceptive use.
- Since independence, average life expectancy has increased substantially
- Infant mortality rate has decreased over the years since independence
- India is facing increased rate of HIV and other sexually transmitted disease cases like other developing countries

All the features described above add different degrees of complexity to the population structure of India and thereby make the prediction task more convoluted. Conventional statistical procedures cannot take care of the chaos contributed by the aforesaid factors. Thus, some methodology is essential that can deal with the non-linearity and chaos intrinsic in Indian population. To the best of the knowledge of the authors of this paper no significant work is available on predicting Indian population through modern mathematical methodology like ANN, Fuzzy Logic etc. Present paper deviates a bit from the conventional procedures and adopts ANN as the research methodology to forecast Indian population.

In the present paper, data are collected from "International Brief: Population trends in India" published by U.S. Department of Commerce Economics and Statistics Administration, Bureau of the Census (Report no IB/97-1). All the population data are available in the scale of thousands and life expectancy data are available in the scale of years.

To frame the input matrix for the ANN, the population are divided into age groups 0 to 14, 6 to 12, 13 to 18, 15 to 44, 15 to 49, 15 to 64, and 65 and over. Instead of dividing the whole population into disjoint subclasses, intersecting subclasses of age are considered to make the data set more suitable for predicting a chaotic system. Male and female population (in thousands) in each subclass, married female population (in thousands) in each subclass, average life expectancy of male, average life expectancy of women, and total fertility rate are considered for the years 1995 and 1997. Purpose is to predict the population (male and female) (in each age group) in the year 2000 on the basis of the data pertaining to 1995 and 1997. From the whole dataset, 75% are considered as training set and 25% are considered as test set. To avoid the sigmoid effect, the data points are transformed into small numerical values through the transformation:

$$x_i = 0.1 + 0.8 \{(x_i - x_{min})/(x_{max} - x_{min})\} \quad \ldots \quad \ldots \quad (7)$$

Initial weight matrix is framed with arbitrary values between –0.5 and +0.5. Now, a non-linear ANN in the form of Multilayer Perceptron is generated with Generalized Delta Learning. The learning procedure is done using the equations (1) through (6). Positive constant $\mu$, called the learning rate is taken as 0.9. After training and testing, correlation coefficients between actual and predicted population of the year 2000 are found to be very high. The overall prediction errors are calculated according to

$$PE = <|x_{actual} - x_{predicted}|>/<x_{actual}> \quad \ldots \quad \ldots \quad (8)$$

Where, <> implies the average over the whole test or training set. When, PE is found to be much smaller than 1, the predictive model is found to be adroit. The correlation between actual and prediction and the relevant prediction errors are presented in Fig.01.

A line diagram (Fig.02) is presented to show the association between actual population in 2000 and the predicted population through non-linear ANN.

## Results and Conclusion

It is apparent from Fig.01 that correlation between actual and predicted population in 2000 is very high both in training and the test phase. Moreover, prediction errors are found to be very less than 1 both in training and test phase. In addition, the error is almost equal to zero in the test phase. These give a qualitative support to the non-linear ANN as a predictive model for population in India. In Fig.02, the actual and predicted populations for both sexes are presented. A close association between actual and predicted data is apparent. Furthermore, correlation between actual and prediction for male is 0.94 and that for female is 0.98. From these correlation values it can be said that, the result is more accurate for female population, from which, it can be inferred that male population is more chaotic.

Thus, the paper establishes suitability of non-linear ANN as predictive tool for population in India. Furthermore, male population in India is found to be more chaotic than female population.

**Acknowledgement**

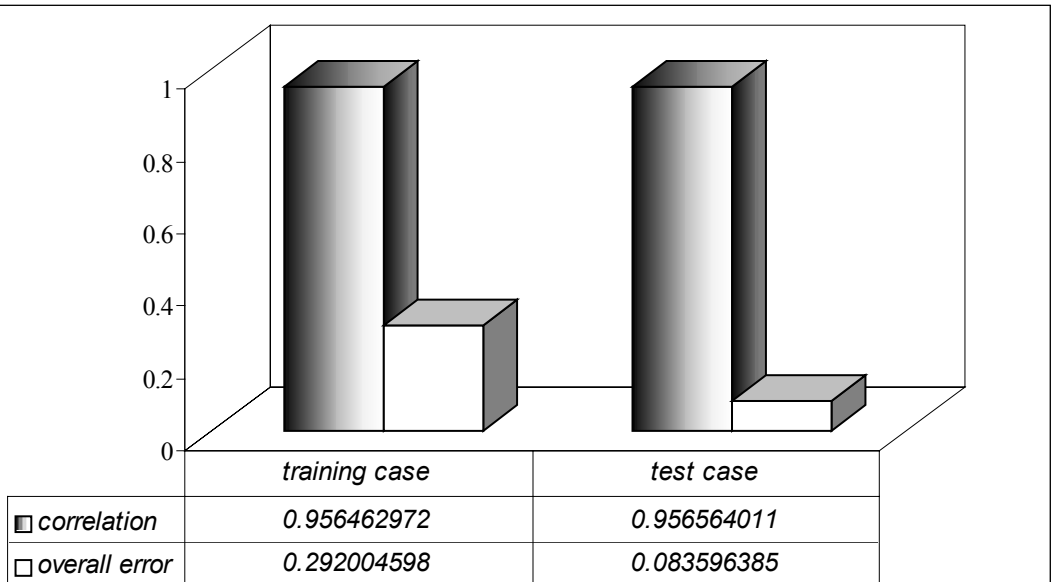

| | training case | test case |
|---|---|---|
| correlation | 0.956462972 | 0.956564011 |
| overall error | 0.292004598 | 0.083596385 |

Fig.01-Graphical presentation of the overall prediction error and correlation between actual and prediction.

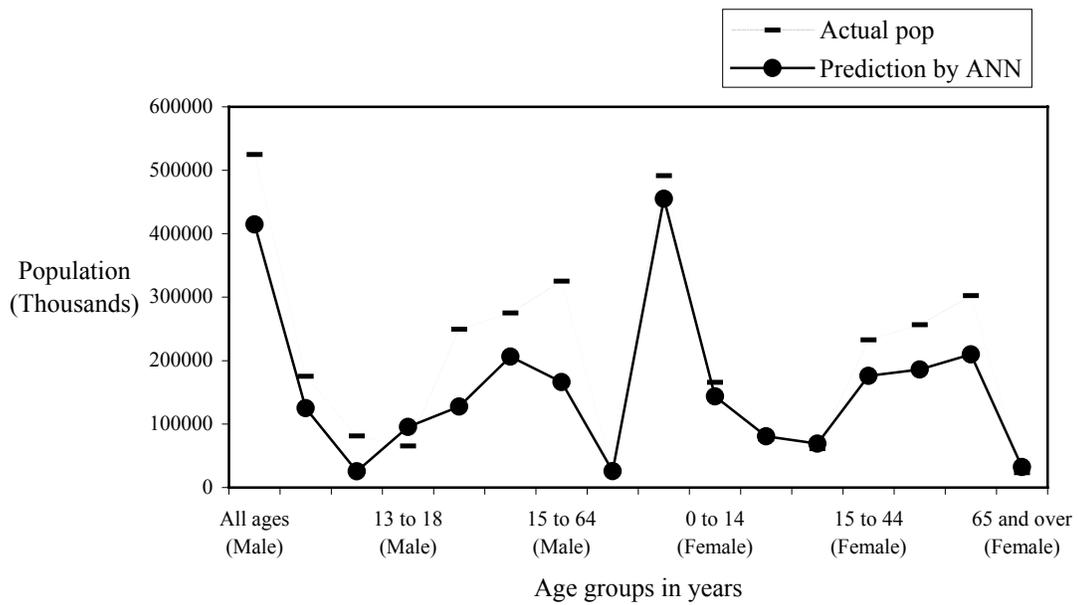

Fig.02-Graphical presentation of the actual population of 2000 and the predicted by three layer non-linear ANN.